\documentclass[prl,aps,twocolumn,showpacs]{revtex4}

\def \beq{\begin{equation}}         \def \eeq{\end{equation}}
\def \beqa{\begin{eqnarray}}        \def \eeqa{\end{eqnarray}}
\def \bea{\begin{array}}        \def \eea{\end{array}}

\def\bio#1#2#3{{Biophys. J. }{\bf #1}, #2 (#3)}

\def\jcp#1#2#3{{J. Chem. Phys. }{\bf #1}, #2 (#3)}

\def\nat#1#2#3{{Nature (London) }{\bf #1}, #2 (#3)}

\def\pnas#1#2#3{{Proc. Natl. Acad. Sci. USA }{\bf #1}, #2 (#3)}
\def\pre#1#2#3{{Phys. Rev. E }{\bf #1}, #2 (#3)}

\def\sci#1#2#3{{Science }{\bf #1}, #2 (#3)}

\usepackage{amsmath}
\usepackage{epsfig}
\begin{document}

\title{Dynamic disorder in receptor-ligand forced dissociation experiments}
\author{Fei Liu}
\email[Email address:]{liufei@tsinghua.edu.cn} \affiliation{Center
for Advanced Study, Tsinghua University, Beijing, 100084, China}
\author{Zhong-can Ou-Yang}
\affiliation{Institute of Theoretical Physics, The Chinese Academy
of Sciences, P.O.Box 2735 Beijing 100080, China}
\author{Mitsumasa Iwamoto}
\affiliation{Department of Physical Electronics, Tokyo Institute
of Technology, 2-12-1 O-okayama, Meguro-ku, Tokyo 152-8552, Japan}
\date{\today}

\begin{abstract}
Recently experiments showed that some biological noncovalent bonds
increase their lifetimes when they are stretched by an external
force, and their lifetimes will decrease when the force increases
further. Several specific quantitative models have been proposed
to explain the intriguing transitions from the ``catch-bond" to
the ``slip-bond". Different from the previous efforts, in this
work we propose that the dynamic disorder of the force-dependent
dissociation rate can account for the counterintuitive behaviors
of the bonds. A Gaussian stochastic rate model is used to
quantitatively describe the transitions observed recently in the
single bond P-selctin glycoprotein ligand 1(PSGL-1)$-$P-selectin
force rupture experiment [Marshall, {\it et al.}, (2003) Nature
{\bf 423}, 190-193]. Our model agrees well to the experimental
data. We conclude that the catch bonds could arise from the
stronger positive correlation between the height of the intrinsic
energy barrier and the distance from the bound state to the
barrier; classical pathway scenario or {\it a priori} catch bond
assumption is not essential.
\end{abstract}

 \maketitle

Since Bell firstly proposed the famous force induced dissociation
rate ~\cite{Bell},
\begin{eqnarray}
k_{\rm off}=k^0_{\rm off}\exp[\beta fx^{\ddag}], \label{Bellexp}
\end{eqnarray}
where $k^0_{\rm off}=k_0\exp(-\beta \Delta G^\ddag)$ is the
intrinsic rate constant in the absence of force, $\Delta G^\ddag$
is the height of the intrinsic energy barrier, $x^{\ddag}$ is
projection of the distance from the bound state to the energy
barrier along the applied external force $f$, and
$\beta^{-1}=k_{\rm B}T$ with $k_{\rm B}$ the Boltzmann's constant
and $T$ absolute temperature, the expression has been demonstrated
experimentally~\cite{Alon,Chen} and widely employed in various
forced dissociation experiments. Later, at least four other models
have been put forward to explain and understand newcome forced
dissociation experiments~\cite{Dembo,Dembo94,Evans97}. In
particular, Dembo proposed a Hookean spring
model~\cite{Dembo,Dembo94} to describe force responses of
receptor-ligand bonds. In addition to predicting that the
dissociation rates of the bonds increase exponentially with the
square of the force, the most important contribution of the model
may be the finding of a ``catch bond" which is defined as
increasing its lifetime when the bond is stretched by the force.
Correspondingly, a bond described by Bell expression is defined
``slip bond" for its lifetime decreases when the force is
applying.

Until recently, the catch bond predicted mathematically was
demonstrated in some biological adhesive bonds which include the
lectin-like bacterial adhesion protein FimH~\cite{Thomas},
P-selctin glycoprotein ligand 1(PSGL-1)$-$P- or
L-selectins~\cite{Marshall,Sarangapani} complex. However, these
experiments also observed that the catch bonds always transit into
slip bonds when the stretching force increases beyond a certain
values, i.e., their lifetimes are shortened again. The
counterintuitive catch-to-slip transition has attracted
considerable attention from experimenters and theorists. Several
kinetic models have been proposed to explain the intriguing
observations in qualitative~\cite{Sarangapani} and quantitative
approaches~\cite{Evans2004,Barsegov,Pereverzev}. We know that the
interface between the ligand and receptor in the adhesive complex
has been reported to be broad and shallow, such as the crystal
structure of PSGL-1$-$P-selectin complex revealed~\cite{Somers}.
In addition, as one type of noncovalent bonds, the interactions
between the molecules are weaker. Therefore it is plausible that
the height and position of the energy barrier of the complex
fluctuate with time due to thermal motion of the whole
macromolecular structure. Dissociation reactions with fluctuating
energy barriers have been studied in terms of rate processes with
dynamic disorder~\cite{Zwanzig}, which was first proposed and
theoretically investigated by Agmon and Hopfield~\cite{Agmon}.
Hence it is of interest to determine whether the fluctuation of
the height and position of the energy barrier induces the
catch-to-slip transition. On the other hand, we also note that in
Bell's initial work and in the other models developed later, the
intrinsic rate constant $k_0$ and the distance $x^{\ddag}$ were
determined and time-independent. It is possible to derive novel
results from relaxation of this restriction. Stimulated by the two
considerations, in the present work we propose a stochastic
Gaussian rate model to quantitatively describe the catch-to-slip
bond transitions. In addition to well predicting the experimental
data, our model provides a new possible physical origin of the
catch bonds: they are likely to be induced by stronger positive
correlation between the fluctuating height of the energy barrier
$\Delta G^\ddag$ and the distance $x^{\ddag}$.

Consider a simple molecular dissociation process under a constant
force $f$,
\begin{eqnarray}
\text{Binding state (B)}\xrightarrow{{k_f} }{\text{Unbinding state
(U)}},
\end{eqnarray}
where the time-dependent forced dissociation rate $k_f(t)$ is a
stochastic variable. If the survival probability $P(t)$ of the
state B is assumed to satisfy the first order decay rate equation
then its formal solution is given by
\begin{eqnarray}
\label{probapprox}
P(t)= \left\langle
{\exp(-\int_0^tk_f(\tau)d\tau)} \right\rangle.
\end{eqnarray}
Cumulant expansion of the above equation~\cite{Kampen} leads to,
\begin{eqnarray}
P(t)\approx\exp\left[-\int_0^t d\tau\langle k_f(\tau)\rangle
+{\text{higher orders terms}}\right],
\end{eqnarray}
where the higher order terms mean the higher correlation functions
of the rate $k(t)$, e.g., the double integral of the second order
correlation function $\langle k_f(t_1)k_f(t_2)\rangle-\langle
k_f(t_1)\rangle\langle k_f(t_2)\rangle$ etc.

According to standard Arrhenius form, we rewrite the Bell
expression as $\langle k_f(t)\rangle=k_o\langle \exp[-\beta
(\Delta G^{\ddag}(t)-fx^{\ddag}(t))]\rangle$. Now the
characteristic of the Bell expression is determined by the two
stochastic processes $\Delta G^{\ddag}(t)$, the energy barrier
height and the distance from the bound state to the barrier
$x^{\ddag}(t)$. The simplest stochastic properties of them are
listed below:
\begin{eqnarray}
&&\langle \Delta G^{\ddag}(t)\rangle=\Delta G^{\ddag}_0\nonumber  \\
&&\langle x^{\ddag}(t)\rangle=x^{\ddag}_0\nonumber  \\
&&\langle x^{\ddag}(t)x^{\ddag}(0)\rangle- \langle
x^{\ddag}(0)\rangle^2=K_{x}(t)\\
&&\langle \Delta G^{\ddag}(t)\Delta G^{\ddag}(0)\rangle-
\langle \Delta G^{\ddag}(0)\rangle^2 =K_{g}(t)\nonumber  \\
&&\langle \Delta G^{\ddag}(t)x^{\ddag}(0)\rangle- \langle \Delta
G^{\ddag}(0)\rangle \langle
x^{\ddag}(0)\rangle=K_{gx}(t)\nonumber.
\end{eqnarray}
Here a stationary process and finite time correlation functions
are assumed. Then using cumulant expansion of $\langle
k_f(t)\rangle$ again and truncating it to second order, we have
\begin{eqnarray}
\label{rateapprox}
\langle k_f(t)\rangle &=& k_0 \exp\left[ -\beta
\Delta G^{\ddag}_0+\frac{\beta^2}{2}K_{g}-\frac{(x^{\ddag}_0-\beta
K_{gx})^2}{2K_{x}}\right]\nonumber \\
&&\times \exp\left[\frac{\beta^2K_x}{2}\left(f-\frac{\beta
K_{gx}-x^{\ddag}_0}{\beta K_x}\right)^2\right], \label{avgrate}
\end{eqnarray}
where $K_{g}$, $K_{x}$, and $K_{gx}$ are the variance and
covariance of and between the two stochastic variables at the same
time point. Because the above consideration has a similar spirit
with the Kubo-Anderson's stochastic line-shape theory~\cite{Kubo},
and the second order truncations in variables $\Delta G^\ddag$ and
$x^\ddag$ are used, we name it Gaussian stochastic rate model
(GSRM).

The average dissociation rate Eq.~(\ref{avgrate}) is so simple
that we can immediately distinguish four different physical
situations according to the definitions of the parameters: (i) if
all $K_x$, $K_g$, and $K_{gx}$ vanish, then average rate is just
the classical Bell expression Eq.~(\ref{Bellexp}); (ii) if both
$K_x$ and $K_{gx}$ vanish or in the absence of the fluctuation of
the distance $x^{\ddag}(t)$, $\langle k_f(t)\rangle $ still keeps
the Bell formula except that the intrinsic dissociation rate
changes into $k_0\exp(-\beta \Delta G^\ddag+\beta^2K_g/2)$,
$i.e.$, the fluctuation of the barrier height speeds up the
dissociation process~\cite{Agmon}. (iii) If both $K_g$ and
$K_{gx}$ vanish or in the absence of the fluctuation of $\Delta
G^\ddag$, then we have
\begin{eqnarray}
\langle k_f(t)\rangle=k_0\exp(-\beta \Delta G^\ddag)\exp\left
[\beta x_0^\ddag f+\frac{\beta^2 K_x}{2}f^2\right ].
\label{squarforce}
\end{eqnarray}
Different from the Bell expression, when the force is larger, the
dissociate rate increases exponentially with the square of force.
This conclusion is very similar with that of Dembo {\it et
al.}~\cite{Dembo}. However the physical origin is completely
different: the square of the force here arises from the
fluctuation of the distance $x^\ddag$. It also means the bond is
still slip and the lifetime of the bond is shorter than that
predicted by the Bell formula. Of course, if the variance of the
distance $K_x$ is very small, the modification to Bell expression
can be neglected. Although the result is interesting, in the
following part we only focus on the fourth case, in which (iv)
both the distance and the barrier height are stochastic variables.
Because the force is positive at the beginning, if $K_{gx}\le0$ or
$(\beta K_{gx}-x^{\ddag}_0)\le0$, the behavior of the average
dissociation rate is then similar to the case in (iii). However if
$x^\ddag_e\equiv\beta K_{gx}-x^{\ddag}_0>0$, we see that the rate
first decreases with the increasing of the force, and then
increases when the force is beyond $f_c\equiv x^\ddag_e/\beta
K_x$, where the new parameters $x^\ddag_e$ and $f_c$ are defined
for they have same dimensions of Distance and Force. Hence our
model predicts the possibility of the catch-slip bond transition
at some critical transition force $f_c$.

\begin{figure}[htpb]
\begin{center}
\includegraphics[width=0.9\columnwidth]{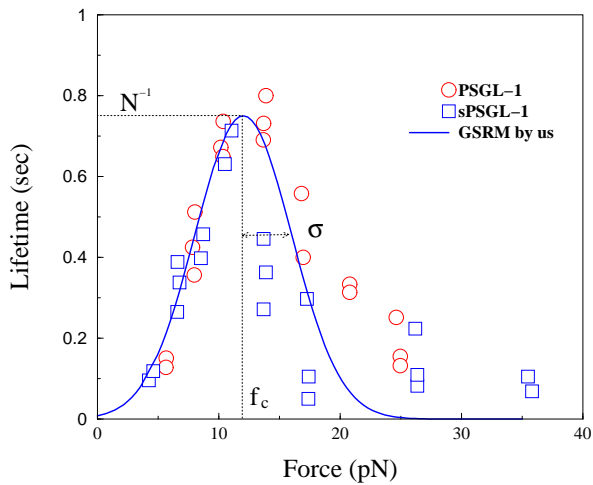}
\caption{(Color online) Average lifetime as a function of the
applied constant force for bonds of dimeric P-selectin with
monomeric sPSGL-1 (blue square symbols)~\cite{Marshall} and the
rescaled dimeric PSGL-1 (red circle symbols) from
Ref.~\cite{Pereverzev}. The blue solid line is given by GSRM. The
definitions of the symbols $N$, $\sigma$, and $f_c$ see
Eq.~(\ref{avgtime}).} \label{figure1}
\end{center}
\end{figure}

We first consider the single molecule constant force rupture
experiment~\cite{Marshall}, where the average lifetime of the bond
sPSGL-1$-$P-selectin was measured. Because the average
dissociation rate $k_f(t)$ is time independent, the survival
probability $P(t)$ is a simple exponential function
$\exp[-t\langle k_f\rangle]$. Following the general definition,
the average lifetime of the bond is just $\bar t(f)=1/\langle
k_f\rangle$. There are six parameters in this model. But in fact
we can combine them into only three: the intrinsic dissociation
rate $k_0^d=k_0\exp{\left[ -\beta \Delta
G^{\ddag}_0+\frac{\beta^2}{2}K_{g}\right]}$, which is the
parameter that experiments can measure in practice, and the
effective distance $x^\ddag_e$ defined above, and $K_{x}$; they
are independent of each other. The average time then is
\begin{eqnarray}
\bar t(f) &=&N^{-1}\exp\left[-\frac{(f-f_c)^2}{2\sigma^{2}}\right] \nonumber \\
&=&\left\{ {k_0^d \exp \left[ { - \frac{{x^{\ddag}_e}^2 } {{2K_x }}} \right]} \right\}^{-1} \\
&&\times\exp \left [ { - \left( {f - \frac{x^\ddag _e } {{\beta
K_x}}} \right)^2 /2(\beta^{-2} K_x^{-1}) } \right ].\nonumber
\label{avgtime}
\end{eqnarray}
It is unexpected to find that the average lifetime of the bond is
a Gaussian-like function with respect to the force: the mean value
is $f_c$, the variance $\sigma$ and a prefactor $N^{-1}$; their
corresponding definitions see the above equation. Apparently they
are still independent. According to the characteristic of Gaussian
function, we can easily estimate the relevant parameters from the
experimental data even without numerical methods; see
Fig.~\ref{figure1}: they are respectively $k_0^d\approx 133.0$~/s,
$x^\ddag_e\approx 2.88$~nm, and $K_x\approx 1$~nm$^2$, whereas the
important correlation coefficient $K_{gx}\ge 12.0$~${\rm pN}\cdot
{\rm nm}^2$ and the catch-slip transition force $f_c\approx 12$ pN
which is directly read out from the experimental data. Here the
estimation is performed at room temperature. We see that our
prediction agrees with the data very well, in particular when the
force is lower than $f_c$~\cite{liuf_exp1}. Interestingly, the
above result also shows that, only through forced dissociation
experiment, we cannot isolate the precise information about the
variance of the energy barrier height and the cross variance of
the height and the distance.

More challenging experiment to our theory is force jump-ramp case,
where the force increases linearly in time from a initial jump
force $f_0$, $f=f_0+rt$, and $r$ is ramp rate~\cite{Evans2004}. In
general, the stationary assumption of the forced dissociation
processes should be more reasonable for the constant case.
Considering that the direct extension of the Bell expression to
time-dependent force case still provides insightful
results~\cite{Evans97}, it is of interest to see what we can get
by extending Eq.~(\ref{avgrate}) to force jump-ramp cases. Because
the experimental data is typically presented in terms of the force
histogram, we calculate the rupture force distribution $P(f,f_0)$
according to definition $P(f,f_0)df=-(dS/dt) dt$, which is given
below,
\begin{eqnarray}P(f,f_0) =\frac{N} {r}\exp \left [
{\frac{{(f - f_c )^2}} {{2\sigma ^2 }} -\frac{N}{r}\int_{f_0 }^f
{df'e^{ {\frac{{(f'  - f_c )^2}} {{2\sigma ^2 }}}}} } \right ].
\label{forcedistr}
\end{eqnarray}
We see that the average lifetime can be extracted from the above
equation by setting $f=f_0$, i.e., $\bar t(f)=1/rP(f_0,f_0)$.

\begin{figure}[htpb]
\begin{center}
\includegraphics[width=0.9\columnwidth]{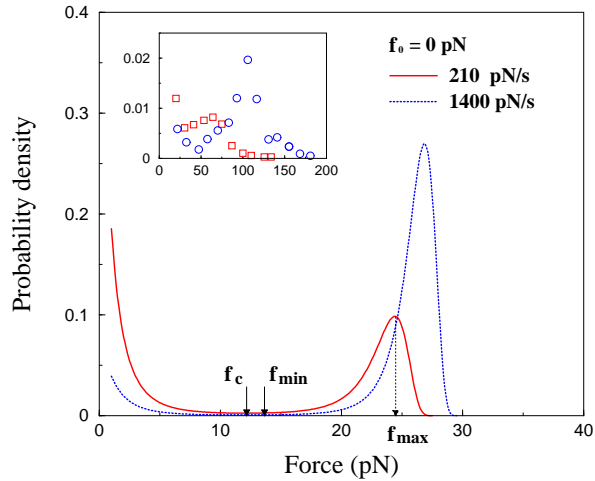}
\caption{(Color online) The distribution of rupture forces under
two loading rates predicted by our model for binding of P-selectin
to sPSGL-1. The symbols of inset are the steady ramp experimental
data~\cite{Evans2004}, where the loading rates are 210 pN/s (red
squares) and 1400 pN/s (blue circles), respectively. We must point
out that the experimental data is for binding of P-selectin to
PSGL-1, while our parameters are for sPSGL-1.}\label{figure2}
\end{center}
\end{figure}

We calculate the force distributions of the steady ramps ($f_0=0$
pN) at ramp rates 210 and 1400 pN/s to compare with the experiment
performed by Evans {\it et al.}~\cite{Evans2004}; see
Fig.~\ref{figure2}, where we use the same parameters obtained
above. We find that the main qualitative characteristics of the
predictions and the experimental data are the same: the
distributions and the force histograms reach the maximum and
minimum at two distinct forces, which are named $f_{\rm min}$ and
$f_{\rm max}$ respectively. This observation could be understood
by setting the derivative of Eq.~(\ref{forcedistr}) with respect
to $f$ equal to zero,
\begin{eqnarray}
f-f_c=\frac{N\sigma^2}{r}\exp\left [\frac{(f-f_c)^2}{2 \sigma^2}
\right ]. \label{extrema}
\end{eqnarray}
Interestingly, Eq.~(\ref{extrema}) has no solutions when the
loading rate is smaller than a critical ramp rate $r_c$, which can
be obtained by simultaneously solving the above equation and its
first derivative
\begin{eqnarray}
1=\frac{N}{r}(f-f_c)\exp\left [\frac{(f-f_c)^2}{2 \sigma^2} \right
].
\end{eqnarray}
Then $r_c=N(f^{*}-f_c)\exp[(f^{*}-f_c)^2/2 \sigma^2]$, here
$f^{*}$ is the force at which the maximum and minimum merge. We
estimate $r_c\approx 9$ pN/s using the current parameters. If
$r\le r_c$, then the distribution is  monotonous and decreasing
function. Another important prediction is that the values of
$f_{\rm min}$ and $f_{\rm max}$ must be larger than the catch-slip
transition force $f_c$. Indeed the experimental observation shows
that the force values at the minimum force histogram are around a
certain values even the ramp rates change 10 folds. (see Figs. 2
and 4 in Ref.~\cite{Evans2004}). According to Eq.~(\ref{extrema}),
when the ramp rate is sufficiently large, we easily obtain
\begin{eqnarray}
f_{\rm min}\approx f_c+\frac{N\sigma}{r}
\end{eqnarray}
by linear expansion. Therefore we predict that $f_{\rm min}$s
observed in Evans {\it et al.} experiment~\cite{Evans2004} are
almost the catch-slip transition force observed in the constant
force rupture experiment performed by Marshall {\it et
al.}~\cite{Marshall,liuf_exp2}. Unfortunately, simply analytic
relationship between $f_{\rm max}$ and $r$ cannot be found from
Eq.~(\ref{extrema}). Even so, the extrema equation implies that
$f_{\rm max}$ is a monotonous and increasing function of the ramp
$r$, but the increasing is very slow and is about $f_{\rm
max}\propto\sqrt{\ln r}$ instead of $f_{\rm max}\propto{\ln
r}$~\cite{Evans2004,Pereverzev}.

We know that the rupture force distribution of simple slip bond
only has a maximum at a certain force value that depends on the
ramp rate~\cite{Izrailev}. Therefor the catch-slip bond can easily
be distinguished from the slip case by the presence of a minimum
on the force rupture density function at nonvanished force.
Because the above analysis is independent of the initial force
jump $f_0$, in order to track the catch behaviors in the force
jump-ramp experiments, the initial force $f_0$ should be smaller
than $f_c$.

In conclusion, we proposed a stochastic dissociation rate model to
explain the intriguing catch-slip bond transitions observed in the
single molecule forced dissociation experiments, while the
fluctuating rate is dependent on the two correlated stochastic
control variables, the energy barrier height $\Delta G^\ddag$ and
the distance $x^\ddag$ between the bound state and the energy
barrier. Compared to the previous models with
five~\cite{Evans2004}, seven~\cite{Barsegov} and four
parameters~\cite{Pereverzev} involved, our model only requires
three physical parameters: $k_0^d$, $x^\ddag_e$, and $K_{x}$.
Moreover, GSRM does not need the classical pathway concept or {\it
a priori} catch bond~\cite{Evans2004,Barsegov,Pereverzev,Bartolo}.
Because there is no direct experiments or molecular structures
supporting the path way scenario, a change in concept would be
important for further experimental study of the catch-slip bonds.

F.L. would like to thank Tokyo Institute of Technology for
hospitality, where this work was completed.


\begin{thebibliography}{99}
\bibitem{liuf_exp1}
Eq.~\ref{avgtime} predicts that the averge lifetime of the bond is
symmetric relative to the critical force $f_c$, while the
experimental data are clearly skewed towards large force. We
simply attribute the discrepancy to the second-order truncation
approximations of the rate in Eq.~(\ref{probapprox}) and the
variables $\Delta G^\ddag$ and $x^\ddag$ in Eq~(\ref{rateapprox}).
Further analysis is needed to give a quantitave conclusion about
the effects of the approximations.
\end{thebibliography}
\end{document}